\documentclass{article}

\usepackage{times}
\usepackage{authblk}
\usepackage{cite} 
\usepackage{url} 
\usepackage[utf8]{inputenc} 
\usepackage{booktabs} 
\usepackage{verbatim}
\usepackage{graphicx}
\usepackage{subfigure}
\usepackage{csquotes}
\usepackage{textcomp}
\usepackage{paralist}
\usepackage{siunitx}
\usepackage{theorem}
\usepackage{hyperref}
\usepackage{amsmath}
\usepackage{cleveref}
\usepackage{xcolor}
\usepackage[colorinlistoftodos]{todonotes}
\usepackage{listings}

\usepackage{color}
\definecolor{mygreen}{rgb}{0,0.6,0}
\definecolor{mygray}{rgb}{0.5,0.5,0.5}
\definecolor{mymauve}{rgb}{0.58,0,0.82}

\newcommand{\footurl}[1]{\footnote{\url{#1}}}

\newcommand{\incubatorrepo}[0]{\footurl{https://github.com/INTO-CPS-Association/example-incubator}}

\newtheorem{assumption}{Assumption}{}

\crefname{assumption}{Assumption}{Assumptions}

\begin{document}

\title{The Incubator Case Study for Digital Twin Engineering}


\author[$\dagger$]{Hao Feng}
\author[$\dagger$]{Cláudio Gomes}
\author[$\dagger$]{Casper Thule}
\author[$\dagger$]{Kenneth Lausdahl}
\author[$\dagger$]{Michael Sandberg}
\author[$\dagger$]{Peter Gorm Larsen}
\affil[$\dagger$]{DIGIT, Department of Electrical and Computer Engineering, Aarhus University, Denmark}

\maketitle
\begin{abstract}  
To demystify the Digital Twin concept, we built a simple yet representative thermal incubator system. 
The incubator is an insulated box fitted with a heatbed, and complete with a software system for communication, a controller, and simulation models. 
We developed two simulation models to predict the temperature inside the incubator, one with two free parameters and one with four free parameters. 
Our experiments showed that the latter model was better at predicting the thermal inertia of the heatbed itself, which makes it more appropriate for further development of the digital twin. 
The hardware and software used in this case study are available open source, providing an accessible platform for those who want to develop and verify their own techniques for digital twins. 

\end{abstract}  

\section{Introduction}

Cyber-Physical Systems (CPSs) are integrations of computation with physical processes. 
Embedded computers act through a network, monitor and control the physical processes \cite{Lee2008}. 

The cyber part refers to the computation part and communications, while the physical part refers to the part of the system interacting with the physical world. 
The environment is the part that may significantly affect the properties of the CPS but cannot directly be controlled. 

An example of a CPS is a car. The physical part in the car contains a body, four wheels, seats, an engine, and so on, which interact with the environment (e.g., wind, road surface). 
The cyber part of the car is different control systems implemented in Electronic Control Units (ECUs) that communicate among themselves through a Controller Area Network. 

Since systems are getting increasingly complex, models are needed to better comprehend their behaviour. 
A model is an abstract representation of a system, built with the goal of understanding one aspect of the system \cite{Kuhne2005a}. 
For example, a Computer-Aided Design (CAD) model of a car may store information about the sizes of the wheels, shapes of the chassis, styles of the car, and so on, but it does not contain information about how the forces acting on the car affect its movement. 
An engineer who desires to analyse the dynamics of the car may build a dynamic model described by the differential equations introduced in, e.g., \cite[Section 5]{Schramm2014}. 

Beyond better understanding, models can assist with optimising the overall system, discovering causes and effects, measuring consequences of changes, and communicating among engineers. 
Once the system is deployed, the models used to build it should not be discarded. Instead, they should be integrated into a Digital Twin (DT).

DT is often defined as a framework that integrates modelling, simulation, monitoring, and optimisation technologies, with the intent of adding value to the use of the Physical Twin (PT) \cite{Tao2019,Fuller2020}.
For instance, a DT provides the ability to run simulations that reproduce and predict the behaviour of the PT under its current operating conditions.
The following important parts are highlighted \cite{Wright2020}: 
\begin{compactdesc}
\item[Data] -- collected from the PT through sensors over time;
\item[Models] -- knowledge about different aspects of both the cyber and the physics of the PT and its environment\footnote{It is worth noting that models can be described at different levels of abstraction, and typically there is a tendency that accurate models also are very slow to analyse for example by means of simulation.}; and
\item[Algorithms] -- techniques that use data and models, manipulating those to generate more data and knowledge (e.g., fault detectors, supervisory controllers, state estimators, optimisers).
\end{compactdesc} 

\begin{figure}[h!]
	\centering
	\includegraphics[scale=0.39]{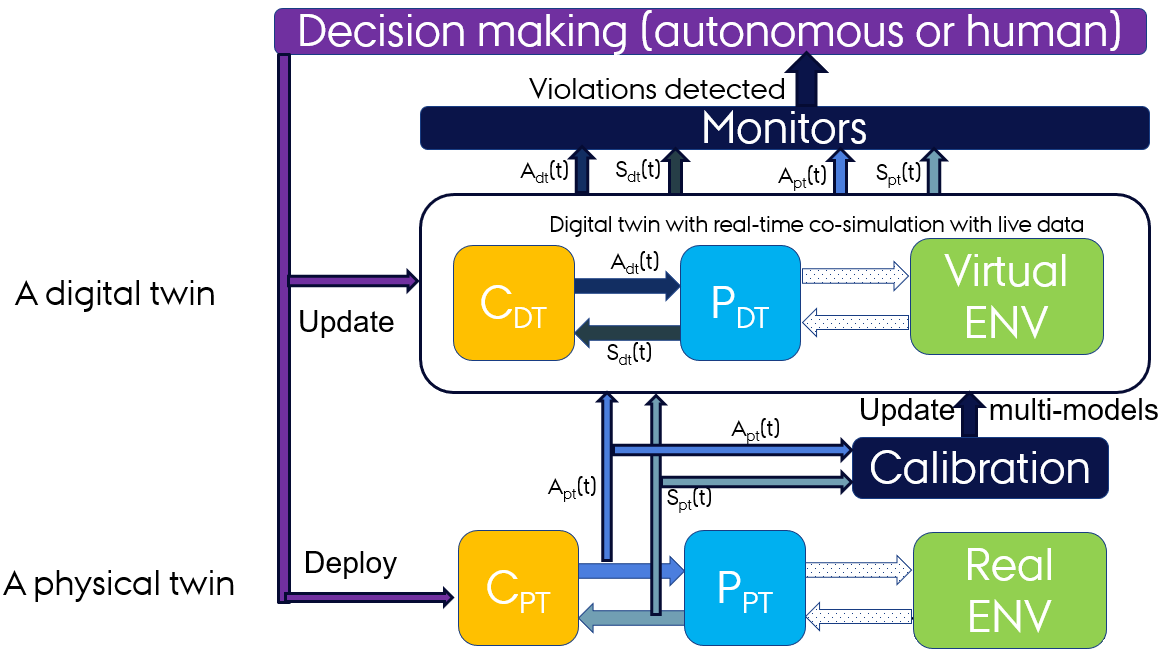}
	\caption{Schematic overview of a CPS-based Digital Twins. The letter \enquote{C} and \enquote{P} represents \enquote{Cyber} and \enquote{Physical} respectively.}
	\label{fig:cpsdt}
\end{figure}

An example common structure of a DT is represented in \cref{fig:cpsdt}.
In \cref{fig:cpsdt}, the DT employs calibration algorithms to update the models of the CPS and environment with the best estimate of the parameters.
The resulting data is then used to inform monitors that check whether the system is performing safely and optimally.
When violations are detected, humans (or other algorithms) may decide to make changes to the physical twin.

DTs increase the value of their physical counterparts by potentially enabling advanced visualisations, reconfigurability (and therefore robustness with respect to changing environment), safety, predictability, and reduced maintenance.

The often vague definitions and claims about DTs makes the concept hard to distinguish from: self-adaptive systems \cite{Weyns2012,Chen2018}, autonomic computing \cite{Kephart2003}, Industry 4.0 \cite{Lasi2014}, models@runtime \cite{Bencomo2014,Bencomo2019}, and supervisory controllers \cite{Karimadini2018}.

We introduce the Incubator case study as an attempt to clearly identify the techniques used in digital twinning.
This system is a simple, yet representative, CPS and PT. 
Detailing how the DT is built is out of the scope of this manuscript, but we sketch how such implementation can be done, and which technologies can be used.
This makes it easier for the reader to see the similarities of DT to the aforementioned related research areas.

The publicly available repository\incubatorrepo\ contains the up-to-date hardware and software specifications of the case study.

The rest of the report is organized as follows: 
\cref{sec:incubator} introduces the incubator system, including the hardware and software setup. 
Then we built two models with calibration method in \cref{sec:modelling}.
\cref{sec:results} shows two kinds of experimental results.
Afterwards \cref{sec:digital_twinning} sketches the goals of the digital twinning and future work. 
Finally, we summarise the report in \cref{sec:summary}. 


\section{Incubator Case Study}\label{sec:incubator}

\begin{figure}[htb]
	\centering
	\includegraphics[scale=0.8]{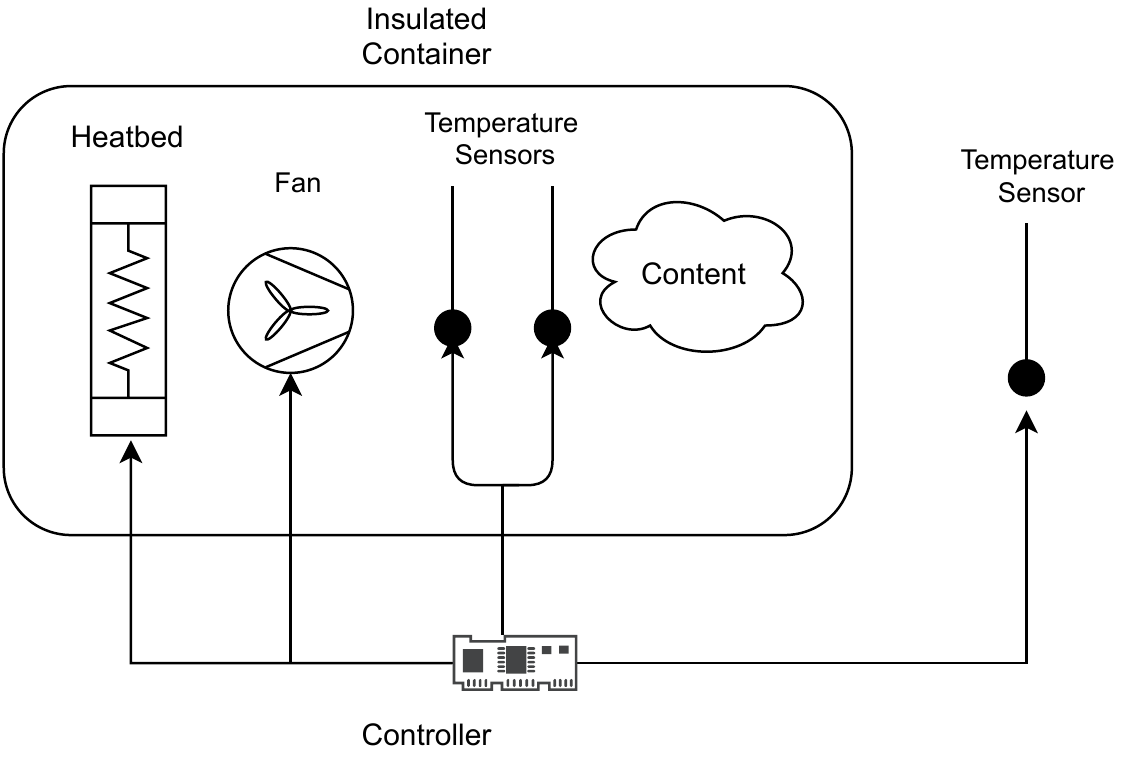}
	\caption{Schematic overview of the Incubator.} 
	\label{fig:incubator}
\end{figure}

The main goal of the incubator is to reach a certain temperature within a box and regulate it regardless of the content inside. 
The physical components in the incubator form a plant that is controlled by a Raspberry Pi.
The overall systematic diagram of the Incubator is shown in \cref{fig:incubator}.

\subsection{Hardware Setup}

We start with a brief description of the hardware components, back-of-the-envelope calculations, and refer the reader to the online repository\incubatorrepo{} for more details.

\begin{figure}[htb]
  \centering
  \subfigure[The styrofoam Box is used as an insulated container.]{\includegraphics[width=.35\linewidth]{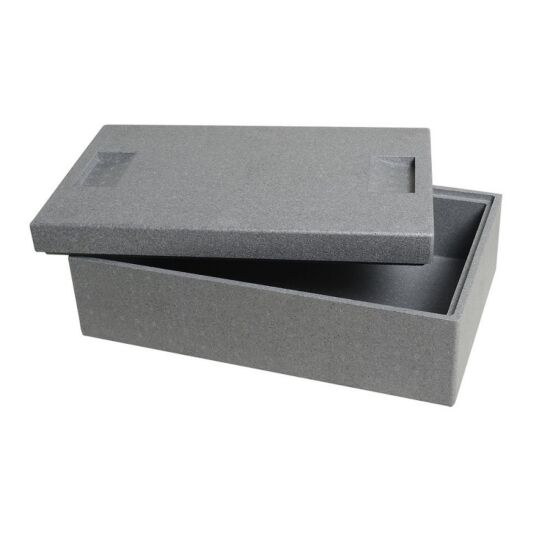}
  \label{fig:styrofoam}}
  \qquad
  \subfigure[The heatbed is approximately $214*214*3$ mm to fit the styrofoam box.]{\includegraphics[width=.35\linewidth]{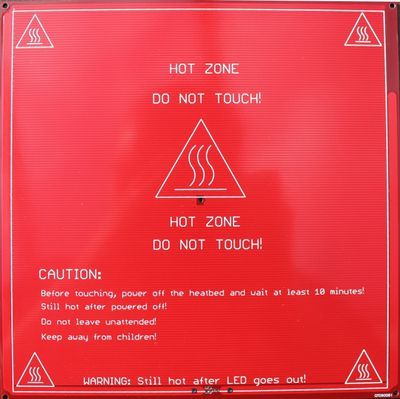}\label{fig:heatedbed}}

  \subfigure[The Fan, used to circulate the air in the styrofoam box. It can be run on 12V or 24V.]{\includegraphics[width=.25\linewidth]{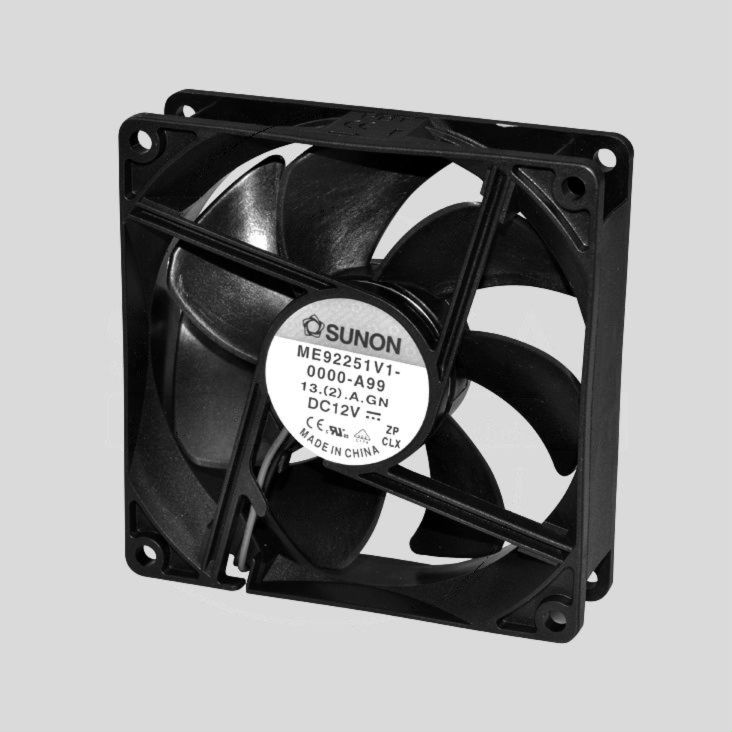}\label{fig:fan}}
  \qquad
  \subfigure[Raspberry Pi 4 Model B]{\includegraphics[width=.25\linewidth]{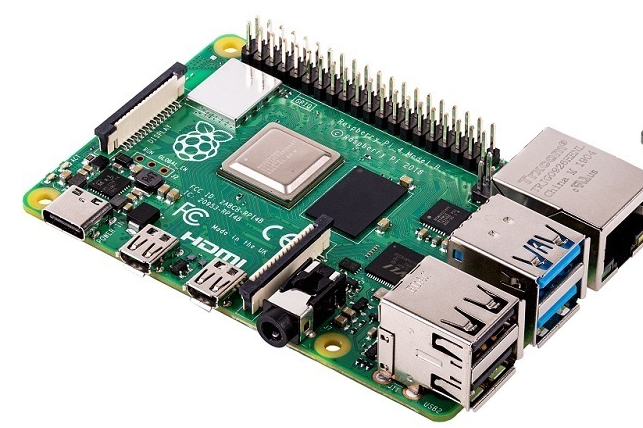}\label{fig:raspberrypi}}
  \qquad
  \subfigure[Temperature Sensor]{\includegraphics[width=.25\linewidth]{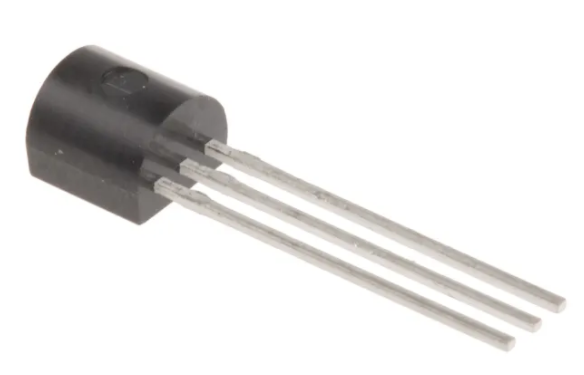}\label{fig:tempsensor}}
  \centering
  \caption{Components used in the incubator}
  \label{fig:components}
\end{figure}

The plant of the Incubator consists of (see \cref{fig:components}):
\begin{compactitem}
    \item A styrofoam box in order to have an insulated container. 
    \item A heatbed to heat up the content inside the styrofoam box.
    \item A fan to distribute the heating inside the box.
    \item Two temperature sensors to monitor the temperature inside the box.
    \item One temperature sensor to monitor the temperature outside the box.
    \item A controller to communicate with the DT, actuate the heatbed, the fan, and read the temperature sensors.
    \item A Printed Circuit Board (PCB) connecting all electrical components.
    \item A power supply.
\end{compactitem}

\subsubsection{The Insulated Container}

The aim of the insulated container (see \cref{fig:styrofoam}) is to preserve the energy generated by the heatbed inside the box as much as possible, so as to minimise the potential impact of the environment around the CPS. 
We chose the styrofoam box (see \cref{fig:styrofoam}) made of polystyrene for its good insulation properties.


The air volume inside the box is about $0.03 m^3$.
This value will be used for estimation of energy and air flow requirements.

\subsubsection{The Heatbed}
\label{sec:heatbed_spec}

The heated content in the incubator is the air inside the box. The heatbed was used (see \cref{fig:heatedbed}) as the surface heat source instead of a bar-like or point-like heat source. 
This is because we want to make the heat energy distributed as uniformly as possible. 

Assuming that
\begin{compactitem}
\item there is no heat transfer into the box walls,
\item the mass of the air inside the box is $0.04 \si{\kilo\gram}$,
\item the air heat capacity is $700 \si{\joule \per \kilo\gram \per \kelvin}$ and it does not change due to the temperatures, pressure and so on,
\end{compactitem}
the heatbed can warm up the air inside the box from $293 \si{\kelvin}$ ($20 \si{\celsius}$) to $300 \si{\kelvin}$ (about $26.85 \si{\celsius}$), by producing 
$$
0.04 \si{\kilo\gram} * (300 \si{\kelvin} - 293 \si{\kelvin}) * 700 \si{\joule \per \kilo\gram \per \kelvin} = 200 \si{\joule}
$$
of energy.

Assuming it delivers at least $100 \si{\watt} = 100 \si{\joule \per \second}$ of power, $200 \si{\joule}$ be provided in under 2 seconds.
As we will show later, this is more than enough power.

\subsubsection{The Fan}

We installed a fan (see \cref{fig:fan}) inside the box to circulate the air inside the box for making the temperature as uniformly distributed as possible. 
The fan airflow is about $0.02 \si{\cubic \meter \per \second}$. 

With a box volume of $0.03 \si{\cubic \meter}$, the fan should take 2 seconds to move all the air within the box.

\subsubsection{The Raspberry Pi}

A Raspberry Pi 4 Model B (see \cref{fig:raspberrypi}) was selected to implement the controller. 
The General Purpose Input/Output (GPIO) connectors on the Raspberry Pi can be used to connect many accessories such as temperature sensors, fan, and the heatbed. Furthermore it supports wireless internet out of the box, with built-in Wi-Fi and Bluetooth, which simplifies deployment of the controller.

\subsubsection{The Temperature Sensor}

It is necessary to get feedback from the temperature sensors for feedback control. To do this, a DS18S20 High-Precision 1-Wire Digital Thermometer (see \cref{fig:tempsensor}) was used for measuring the temperature. 
This temperature sensor can measure temperatures from -55 $\si{\celsius}$ to +125 $\si{\celsius}$ with an accuracy of 0.5 $\si{\celsius}$ from -10 to +85 $\si{\celsius}$, which is suitable for the physical incubator. 
And each DS18S20 has a unique 64-bit serial code, which allows multiple DS18S20s to function on the same 1-Wire bus. 
Such 1-wire bus support to connect multiple sensors with only one GPIO.
In the incubator system, three sensors were deployed for detecting the temperature.

\subsubsection{The PCB Design}

\begin{figure}[h!]
	\centering
	\includegraphics[scale=0.285]{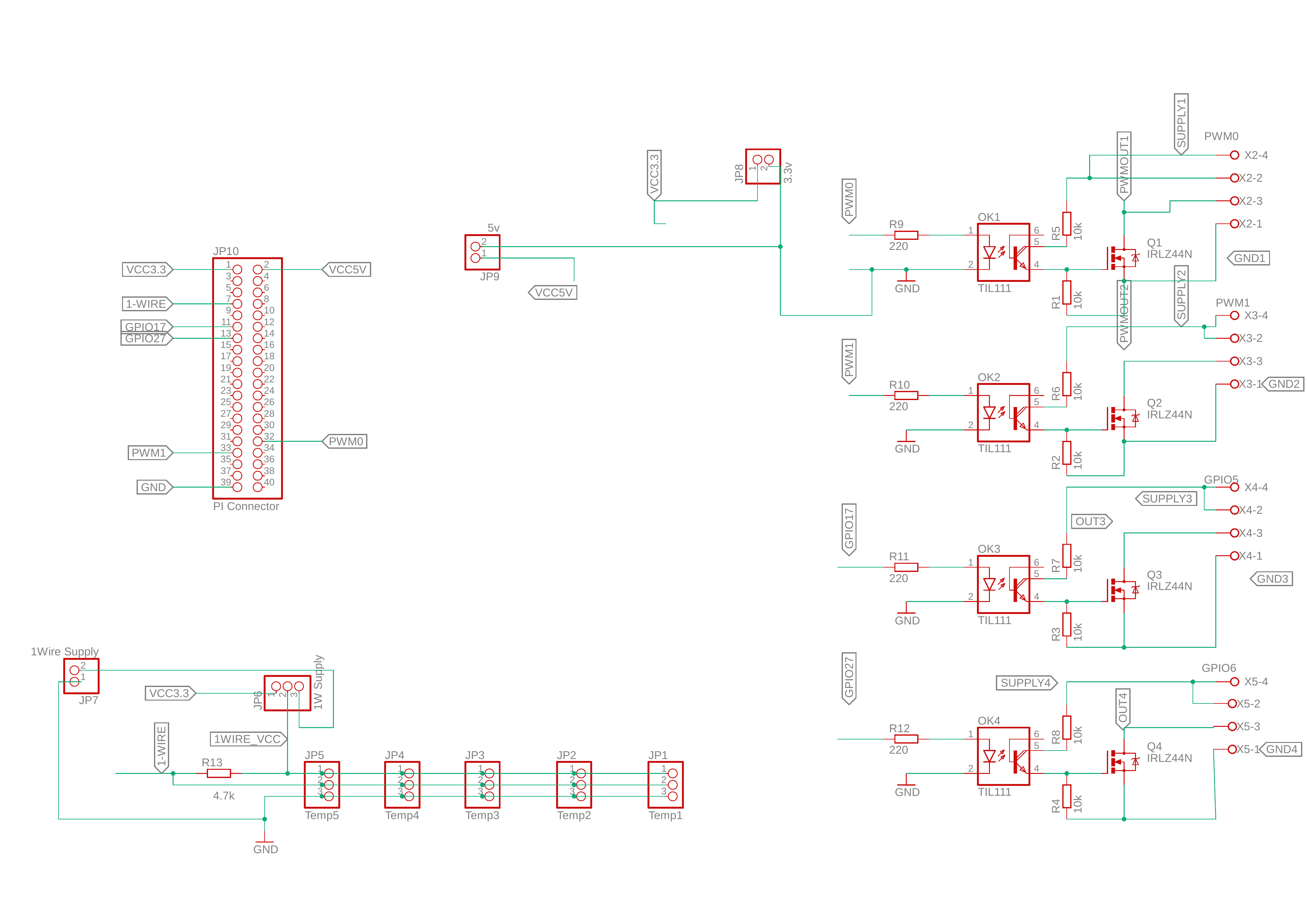}
	\caption{PCB schematics.}
	\label{fig:pcbschematics}
\end{figure}

Controlling the heatbed requires approximately $4 \si{\ampere}$ current which is not supported by the Raspberry Pi. 
Instead, MOSFETs are used to separate the control signals from the Raspberry Pi and the driving currents. 
In order to decouple the control signals from the Raspberry Pi to the MOSFETs, we use optocouplers. 
The PCB schematic is shown in \cref{fig:pcbschematics}.

\subsubsection{The Connections}

\begin{figure}[htb]
  \hfill
  \subfigure[Styrofoam box configuration.]{\includegraphics[trim={0 0.2cm 0 0},clip,width=5cm]{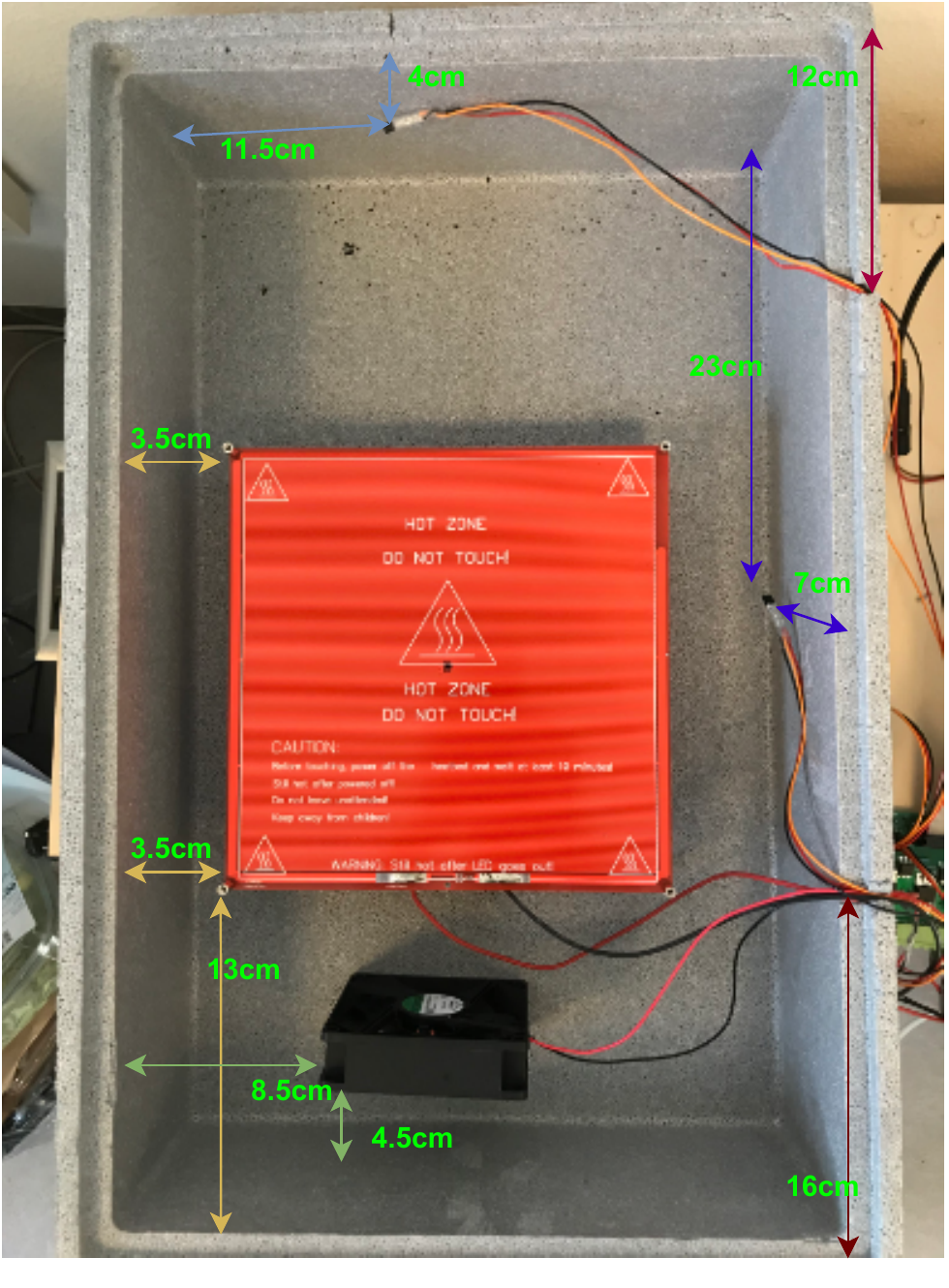}}
  \hfill
  \subfigure[PCB board connection.]{\includegraphics[width=5cm]{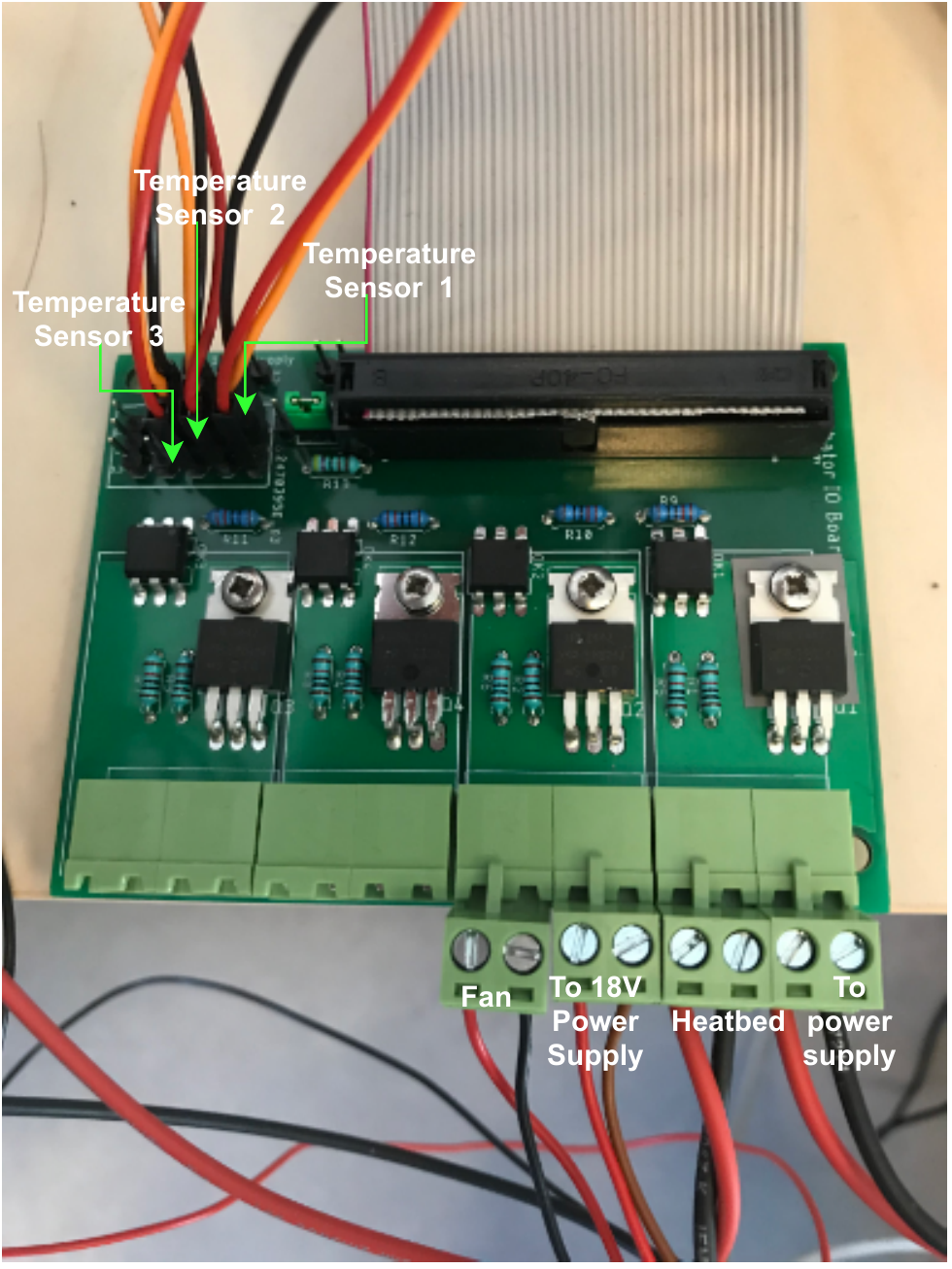}}
  \hfill
  \caption{Connections of components.}
  \label{fig:connectiondetail}
  \end{figure}

The configuration details are shown in \cref{fig:connectiondetail}.

\subsection{The Incubator Software Setup}

\begin{figure}[htb]
	\centering
	\includegraphics[scale=0.55]{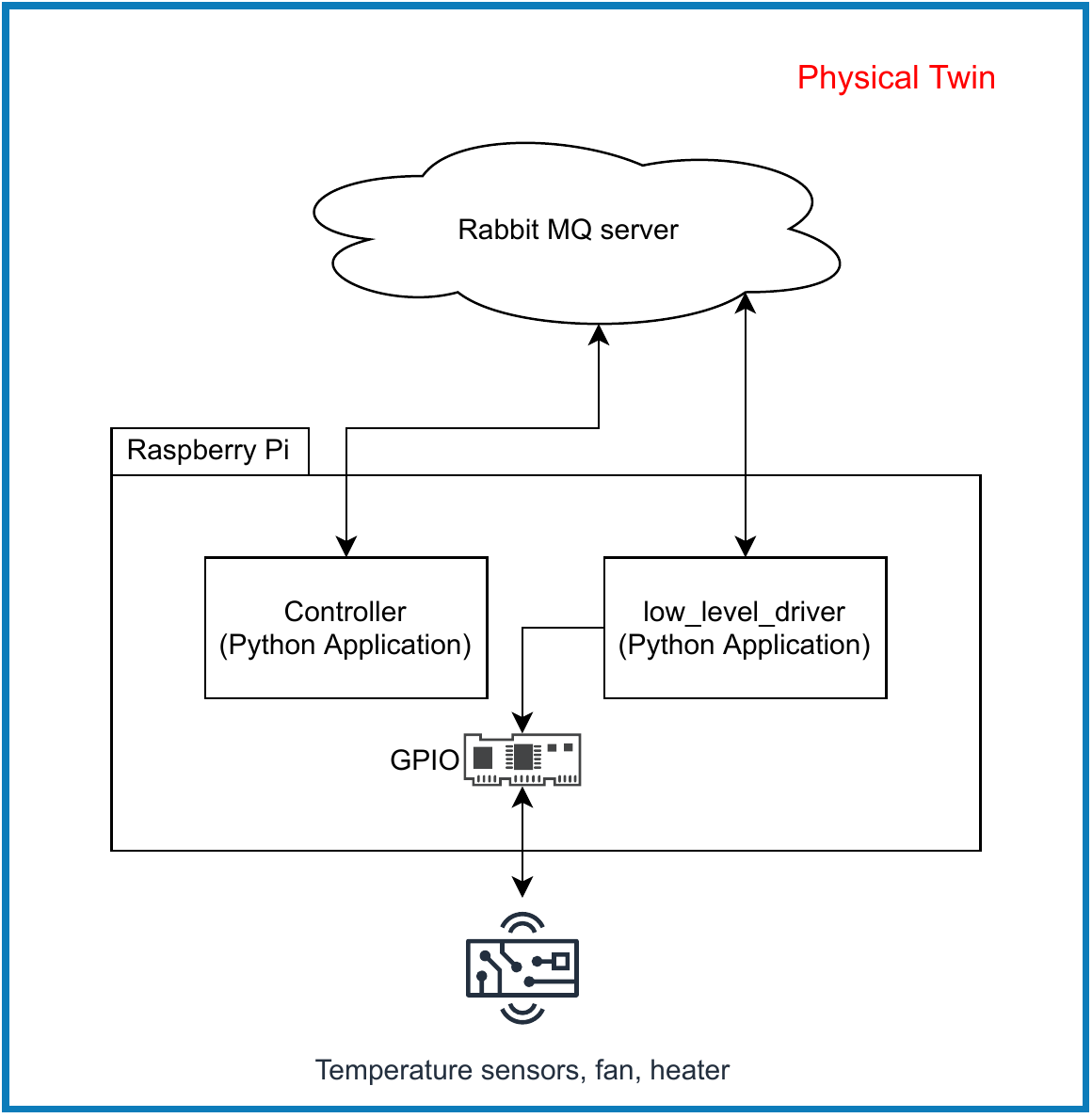}
	\caption{Schematic of the software deployment of the components.}
	\label{fig:softwaresystem}
\end{figure}

The software setup includes a controller, a low-level driver, and a communication server for delivering data. Fig.~\ref{fig:softwaresystem} shows the schematic of the software system.
\begin{compactdesc}
\item[Communication Server] -- We used a Rabbit MQ server, running on the raspberry pi itself.
\item[Low-level driver] -- Abstracts the low level communication with the sensors and actuators.
\item[Controller] -- Implements the control logic required to regulate the temperature inside the incubator. It receives the measured data and sends commands through the communication server. 
\end{compactdesc} 

\subsubsection{The Communication Server}

For the case study, the RabbitMQ\footurl{https://www.rabbitmq.com/} server was deployed for synchronisation and communication. 
With tens of thousands of users, RabbitMQ is one of the most popular open source message brokers. 
RabbitMQ is a mature, lightweight, and easy to deploy on premises and in the cloud. 

The controller on the Raspberry Pi is listening from the RabbitMQ server for commands of controlling the fan and the heatbed and at the same time sharing the control states to the RabbitMQ server. 

\begin{figure}[htb]
	\centering
	\includegraphics[scale=0.5]{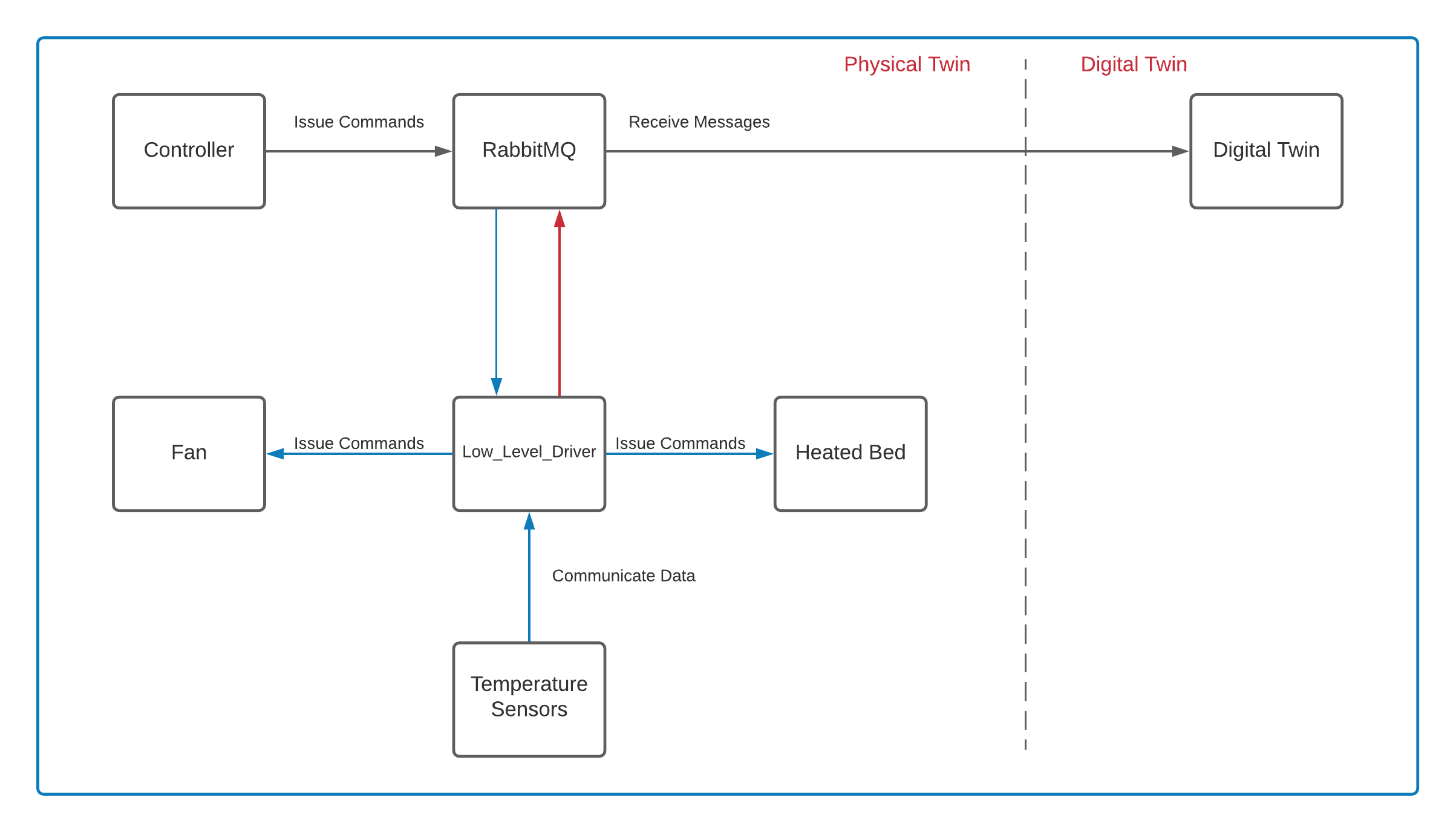}
	\caption{Overview of communication in the Incubator. The arrows represent the information flow.}
	\label{fig:software}
\end{figure}

We choose RabbitMQ because it greatly facilitates the digital twinning process, as the DT can be implemented by components that listen to the messages being passed between the controller and the load level driver, as illustrated in \cref{fig:software}. 

\subsubsection{The Low-Level Driver}

The low-level driver is used to interact with the PT and to issue the states of the PT to the RabbitMQ server. 
In order keep the device in a safe state, a safe protocol is incorporated into the low-level driver. 

The low level driver periodically reads the temperature data, and checks if there are commands from the controller, through the communication server. 
The temperature data is uploaded to the communication server.

\subsubsection{The Controller}

The controller used in the case study is a comparable to a bang-bang controller, which a small difference described below. 
The control strategy is executed after receiving the temperature measurements. 

\begin{figure}[htb]
	\centering
	\includegraphics[width=0.4\textwidth]{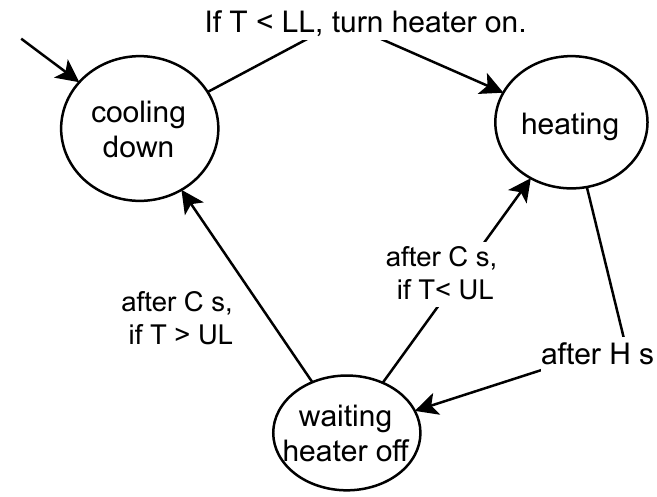}
	\caption{Controller Statechart.}
	\label{fig:controller_statechart}
\end{figure}

The principle of bang-bang control is that when a reference temperature is given to the controller, the controller issues a heating signal turning on the heatbed until the measured temperature from sensors reaches the reference temperature.
Because of the delayed effect of the temperature, the controller needs to wait after each actuation, to make sure the temperature does not rise too much.
\Cref{fig:controller_statechart} shows the state chart of the controller.
The controller parameters are:
\begin{compactdesc}
\item[LL] -- Lower limit for temperature;
\item[UL] -- Upper limit for temperature;
\item[H] -- Heating duration;
\item[C] -- Waiting duration.
\end{compactdesc}

\section{Modelling the Physical Twin} \label{sec:modelling}

The dynamics model we created mainly focuses on predicting the temperature inside the styrofoam box. To simplify the model, a few assumptions were made.

\begin{assumption}\label{ass:temp_uniform}
The temperature inside the box is distributed uniformly.
This was accomplished by deploying a fan inside the box ensuring distribution of the air inside the box.
We test this assumption in \cref{sec:results}.
\end{assumption}

\begin{assumption}\label{ass:box_materials}
The box walls do not accumulate heat, or such heat accumulation has little effect on the air temperature inside the box.
\end{assumption}

\begin{assumption}\label{ass:heat_capacity_const}
The specific heat capacity of the air inside the box is assumed constant, even when pressure and humidity may change.
\end{assumption}

\begin{assumption}\label{ass:heatbed_capacitance_null}
The majority of electrical power is transferred to heat energy, which means the heatbed inside the box does not absorb the heat energy and the heat energy generated by the fan was ignored as well.
\end{assumption}

Two different models were built:
\begin{description}
\item [Model A:]  relies on all assumptions.
\item [Model B:]  relies on all assumptions except \cref{ass:heatbed_capacitance_null}.
\end{description}

To obtain the dynamic models of the temperature, the model was developed by considering the way the energy flows. 
If the amount of the energy used for heating the air is acquired, then this energy can be converted into the temperature inside the box by calculating a basic heat equation: 
\begin{equation}
  Q=cm\Delta T \label{equ:basicheateuqation},
\end{equation}
where $Q$ represents the energy transferred, $m$ equals the mass of the heated object, $c$ is the specific heat capacity which is dependent on many factors such as temperature and pressure, in this project to simplify the dynamic model, $c$ is considered to be constant, and not dependent of temperature and pressure following  \cref{ass:heat_capacity_const}. 
$\Delta T$ is the temperature change either in Kelvin or in \textdegree{}C. 
All the units are SI-units. Based on this equation, the energy can be converted into temperature, which is the basis of the two time-dependent models. 


We now explain the derivation of Model A, and Model B is described in \cref{sec:model2}.

\subsection{Model A}


In model A, only the transformation of the energy between the air inside the box and outside of the box is considered. This means all the energy from the power supply is transferred to the air inside the box (recall \cref{ass:heatbed_capacitance_null} above). 
The total energy brought in by the power supply is:
\begin{equation}
  E_{power\_in} = VI\Delta t \label{equ:powerin},
\end{equation}
where $V$ and $I$ are the voltage and current provided by the power supply and $\Delta t$ is the time interval. 

Since the temperature of the air inside the box differs from that of the box, part of the energy flows from the air inside the box to the box. A parameter $G_{box}$ (unit: \si{\joule \per \kelvin}) was used to link the difference between the room temperature and the air temperature inside the box with the energy flowing to the box, which is:
\begin{equation}
  E_{power\_out} = G_{box}(T_{bair}-T_{room})  \label{equ:powerout},
\end{equation}
where $T_{bair}$ represents the temperature of the air inside the box and $T_{room}$ is the room temperature.

By subtract \cref{equ:powerout} from \cref{equ:powerin}, we can get the difference of the energy. Substitute the difference of the energy to the $Q$ in \cref{equ:basicheateuqation}, a new equation is obtained:
\begin{equation}
  \frac{dT_{bair}}{dt}=\frac{1}{c_{bair}m_{bair}}[VI\Delta t-G_{box}(T_{bair}-T_{room})].
\end{equation}
Since $c_{bair}$, the capacity of the air inside the box, is assumed to be constant but unknown and $m_{bair}$, the mass of the air inside the box, is a constant, a new parameter $C_{air}$ (unit:\si{\joule \per \kelvin}) is used to replace $c_{bair}*m_{bair}$. 
A more compact equation with model A is:
\begin{equation}
  \frac{dT_{bair}}{dt}= \frac{1}{C_{air}}[VI\Delta t-G_{box}(T_{bair}-T_{room})].\label{equ:twoparmodel}
\end{equation}

\subsection{Model B}
\label{sec:model2}

The main difference between Model B and Model A is that Model B relaxes \cref{ass:heatbed_capacitance_null}. 
In other words, it considers the heatbed to also accumulate heat.

Similar to $G_{box}$, $G_{heater}$ (unit:\si{\joule \per \kelvin}) is used to describe the energy transferred from the heatbed to the air. 
The heat bed has its own specific heat capacity and mass. 
The energy transferred from the heatbed to the air is:
\begin{equation}
  E_{heater2air} = G_{heater}(T_{heater}-T_{bair}).
\end{equation}
And the energy retained in the heatbed is: 
\begin{equation}
  E_{heater} = VI\Delta t- E_{heater2air}.
\end{equation}
This energy makes the temperature of the heatbed increase. 

The process of energy flowing from the air to the box is the same as model A. 
The energy used for heating the air can be calculated by: 
\begin{equation}
  E_{air} = G_{heater}(T_{heater}-T_{bair}) - G_{box}(T_{bair}-T_{room}).
\end{equation}
Both the energy $E_{heater}$ and $E_{air}$ are used for heating the heatbed and the air. 
According to \cref{equ:basicheateuqation}, the changes in the temperature of the heatbed and the air are: 
\begin{equation} \label{equ:fourparmodel}
  \begin{split}
    \frac{dT_{heater}}{dt} & = \frac{1}{C_{heater}} (VI\Delta t- G_{heater}(T_{heater}-T_{bair}))\\
    \frac{dT_{bair}}{dt} &=\frac{1}{C_{air}} [G_{heater}(T_{heater}-T_{bair}) - G_{box}(T_{bair}-T_{room})].
  \end{split}
  \end{equation}

The model (\cref{equ:fourparmodel}) considered more detailed than Model A, has four parameters $C_{heater}$, $C_{air}$ (unit:\si{\joule \per \kelvin}), $G_{heater}$, and $G_{box}$ (unit:\si{\joule \per \kelvin  }). 

\subsection{Calibration} \label{sec:calibration}

Although the dynamics models have been built, they contain free parameters. 
In order to utilise the models, the models need to be calibrated for the incubator. 
The results of the calibration are detailed in \cref{sec:calibration_results}, and here we detail the procedure.

A non-linear least squares solver was used to calibrate the parameters of the models derived above.
Then the models are evaluated using the estimated parameters. 
Afterwards the resulting behaviour is compared with a data trajectory that is obtained by running an experiment of the incubator. 
Finally using an optimisation function inside the python Scipy package \cite{SciPy1.0Contributors2020} to obtain the values of the parameters. 
Such a function finds the parameters that minimizes the residual error.
When the residual error is small enough, the optimised parameters can be obtained. 

The principle of the package is based on least-square, it assumes that:
\begin{equation}
  y=f(x,\theta)+\epsilon
\end{equation}
where $f(x,\theta)$ is the model needed to be calibrated, $\theta$ are the parameters represented in a vector in the model, and $y$ is the label or the measured data.

A cost function or an objective function can be provided as:
\begin{equation}
  J=\sum(y_i-f(x_i,\theta))^2
\end{equation}
The objective is to acquire a minimal value of $J$, and at the
same time, the corresponding $\theta$ are the value needed. 
In order to minimise the cost function, the most commonly used one is gradient descent. 
After multiple iterations, the value of the cost function goes to a sufficient small value and the $\theta$ are obtained simultaneously.


\section{Experimental Results} \label{sec:results}

Two kinds of experiments have been conducted. One is for testing \cref{ass:temp_uniform}. Another has been conducted to calibrate the parameters of the models. 
The source code used for the experiments is available online\incubatorrepo{}.	

\subsection{Uniform Temperature Experiments}

\begin{figure}[h!]
	\centering
	\includegraphics[scale=0.5]{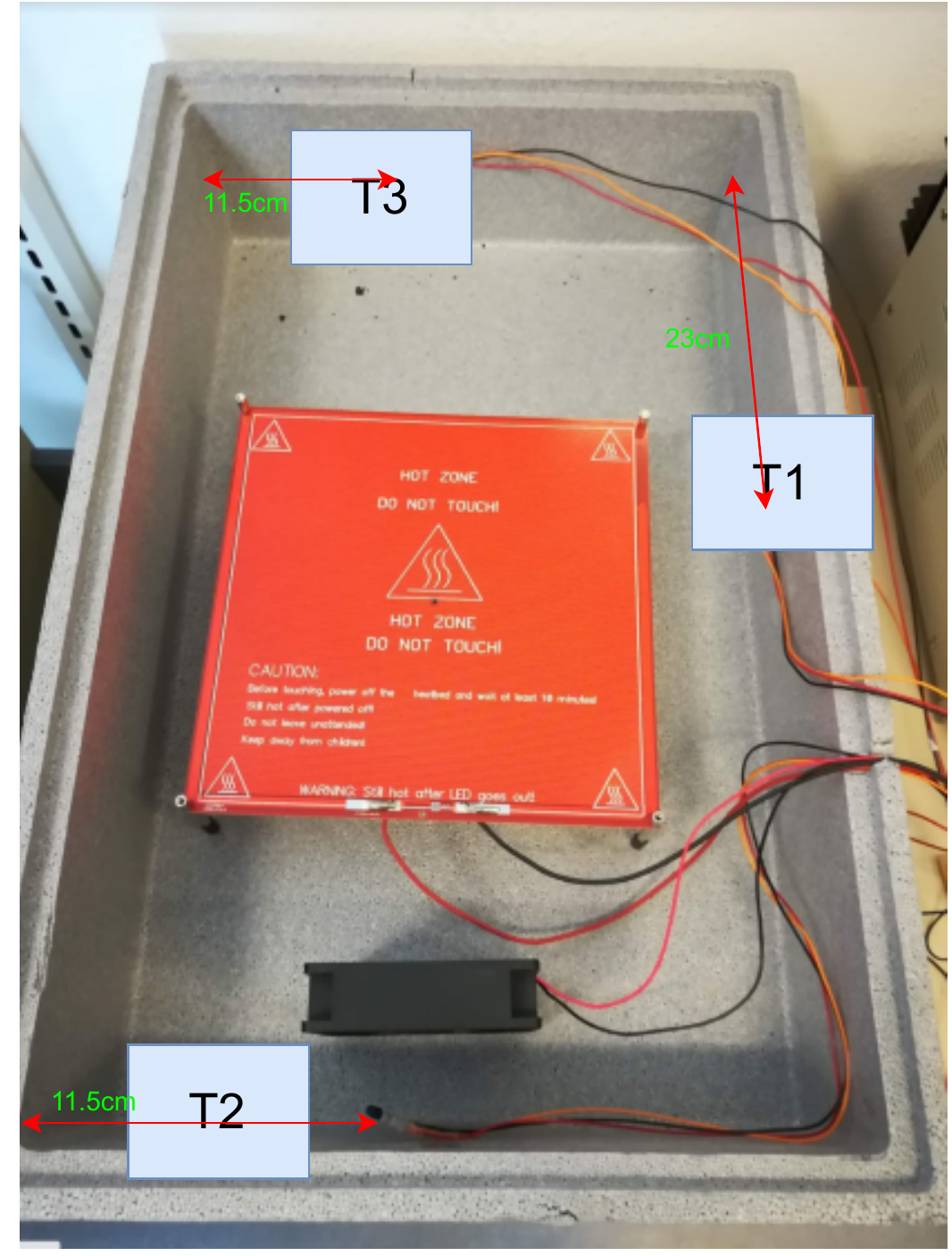}
	\caption{Three sensors are placed in different places.}
	\label{fig:sensorsetup}
\end{figure}

The temperature inside the box is assumed to be distributed uniformly, thanks to the fan. 
An experiment was conducted to test the uniformity of the temperature. 
In this case, it is not necessary to know the temperature of the room, so the sensor measuring the room temperature was used to test the air temperature inside the box. 
The setup with the sensors can be seen in \cref{fig:sensorsetup}. 

\begin{figure}[h!]
	\centering
	\includegraphics[scale=0.23]{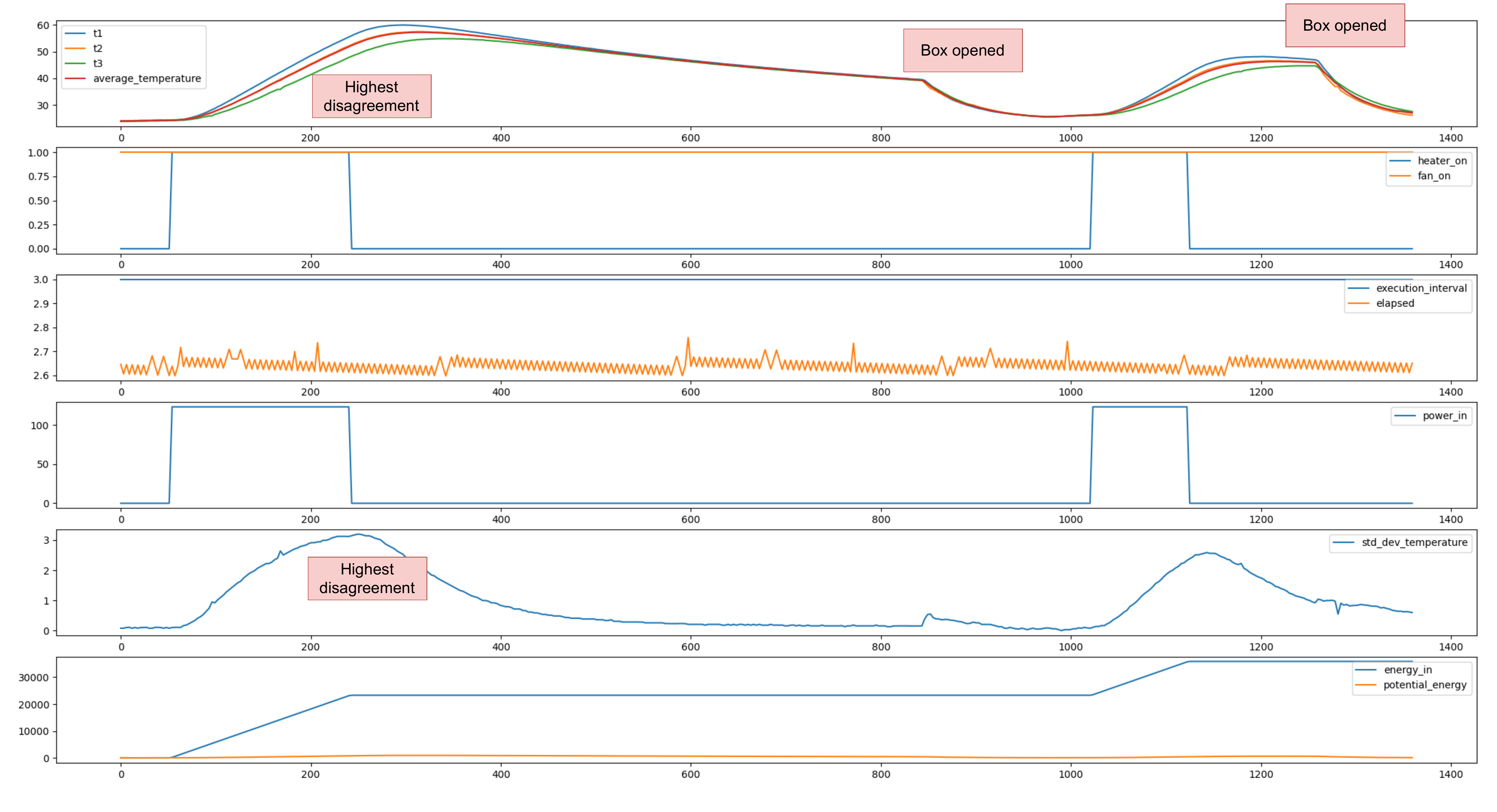}
	\caption{Experimental results of a uniform temperature inside the box. The first subfigure shows the temperatures measured from the three sensors change over time. The temperature measured from the sensor T2 is overlapped by the average temperature, which is hard to read. The other subfigures show different information regarding the experiment such as the status of the heatbed, the power introduced in the system, etc.}
	\label{fig:uniformexperiment}
\end{figure}

During the experiment, the fan is always turned on in order to circulate the air inside the box and the fan blows the wind from the bottom to the top in \cref{fig:sensorsetup}. The heatbed was turned on during two periods. 
The experimental results are shown in \cref{fig:uniformexperiment}.

From the first subfigure in \cref{fig:uniformexperiment}, it seems that the temperature correlates with the distance to the heatbed and the average temperature matches the temperature measured from T2 in \cref{fig:sensorsetup}.  

\begin{figure}[h!]
	\centering
	\includegraphics[scale=0.5]{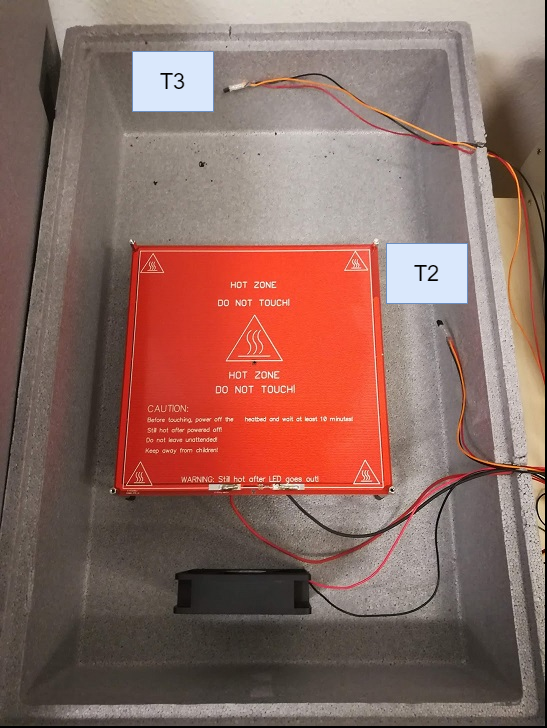}
	\caption{Optimized setup for sensors. The position of T2 and T3 are the same with \cref{fig:sensorsetup}}
	\label{fig:newsensorsetup}
\end{figure}

Since only three temperature sensors are included in the box and one is necessary for measuring the room temperature, sensor T2 in \cref{fig:sensorsetup} was rearranged to the new place in \cref{fig:newsensorsetup} and sensor T1 in \cref{fig:sensorsetup} was taken to be outside of the box measuring the room temperature. 
The new setup of the sensors is shown in \cref{fig:newsensorsetup}. One was placed in the hottest position which is T2 position in \cref{fig:newsensorsetup} and another one was placed in the coldest position, T3 position in \cref{fig:newsensorsetup}. 
Taking the average of those two, the average temperature of the air inside the box is acquired. 

\begin{figure}[h!]
	\centering
	\includegraphics[scale=0.5]{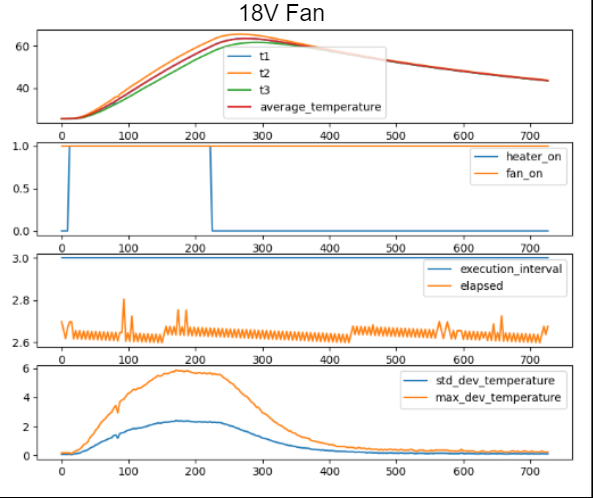}
	\caption{The experimental result with 18 voltage on the fan.}
	\label{fig:18Vfan}
\end{figure}

In order to determine whether a stronger wind flow contributes a more uniform temperature, one additional experiment was conducted based on the setup in \cref{fig:sensorsetup}. 
All the conditions were the same except the voltage of the fan which is 18$V$ while in the previous experiment it was 12$V$. 
The result is shown in \cref{fig:18Vfan}.

The previous setup in \cref{fig:uniformexperiment} showed a discrepancy of a maximum of 7 \textdegree{}C between the three temperature measurements while the maximum discrepancy in a new setup with 18 voltage on the fan is 6 \textdegree{}C.

\subsection{Calibration Experiment} \label{sec:calibration_results}

\begin{figure}[h!]
	\centering
	\includegraphics[scale=0.5]{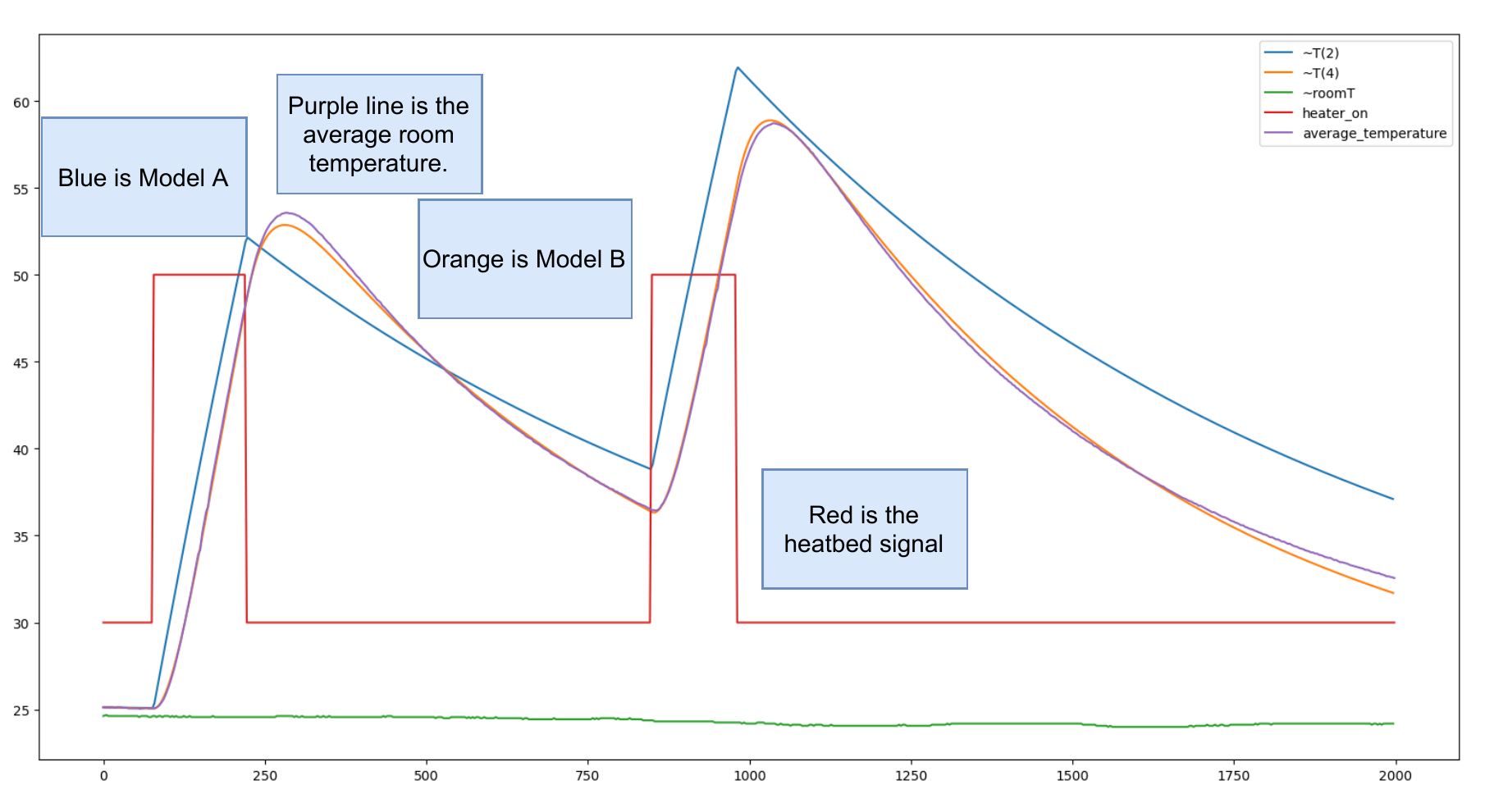}
	\caption{The calibration result.}
	\label{fig:calibrationresults}
\end{figure}

Based on the optimised setup for sensors (\cref{fig:newsensorsetup}), a calibration experiment was conducted and the calibration method has been described in \cref{sec:calibration}. The experiment was used for generating a data trajectory. 
The calibrated parameters for the model A are $C_{air} (unit:\si{\joule \per \kelvin  })=616.56$ and $G_{box} (unit:\si{\joule \per \kelvin  })=0.65$. And the calibrated parameters for the model B are $C_{air} (unit:\si{\joule \per \kelvin  })=486.12$, $G_{box} (unit:\si{\joule \per \kelvin  })=0.856$, $C_{heater} (unit:\si{\joule \per \kelvin  })=33.65$, and $G_{heater} (unit:\si{\joule \per \kelvin  })=0.87$. 
The calibration result can be seen from \cref{fig:calibrationresults}.
Model A is sufficient to get a rough approximation of the system magnitude and time scale, whereas the model B is better at predicting the inertia of the heatbed itself.





\section{Digital Twinning Goals} \label{sec:digital_twinning}

In this section, we sketch the goals of the digital twinning process for the incubator case study.
This constitutes our future work.

Our goal is to implement a DT framework that enables
\begin{compactdesc}
\item[Real-time visualization of the incubator status] -- achieved by listening to the controller and low level driver messages, and relaying that data into a time series database, and plot the data into a dashboard.
\item[On demand calibration] -- achieved by automatically running the calibration process described in \cref{sec:calibration}. This is similar to the tracking simulator implemented in \cite{Legaard2020}.
\item[Anomaly detection and Heatbed state estimation] -- This is achieved by running a Kalman filter \cite{Kalman1960} that uses Model B developed in \cref{sec:modelling} and correlates the measure data to the (non-measured) heatbed temperature.
\item[What-if simulations] -- This is implemented by running co-simulations that relay the historical or real-time data from the PT (e.g., similar to what is done in \cite{Thule2020}). It can be used to, e.g., optimize the controller parameters introduced in \cref{fig:controller_statechart}.
\item[Self-adaptation loop] -- This is the functionality that truly \emph{closes the loop} of the DT. We envision that, whenever a new object (imagine a bucket of cold water) is placed in the incubator, the DT will:
  \begin{compactenum}
    \item detect that the plant behavior has changed (given by the anomalies detected by the Kalman filter);
    \item schedule an experiment to gather relevant data (e.g., let the plant cool down to a safe temperature, then ramp up heating for some time).
    \item configure the controller for the new experiment;
    \item gather the experiment data;
    \item run the calibration of Model B for new experiment;
    \item re-configure Kalman filter with new parameters;
    \item run what-if simulations to optimize the controller behavior; and
    \item finally re-configure the controller.
  \end{compactenum} 
\end{compactdesc}

\section{Summary}\label{sec:summary}

We described the implementation and modelling of the incubator case study.
Hardware, software, and datasets are available online\incubatorrepo{}.
Using this example to set the terminology, we proceeded to demystify the DT concept and its goals.

\newpage
\bibliographystyle{unsrt}
\bibliography{bibliography_generated_hao,bibliography_generated_claudio}

\begin{thebibliography}{10}

\bibitem{Lee2008}
Edward~A. Lee.
\newblock Cyber {{Physical Systems}}: {{Design Challenges}}.
\newblock In {\em 11th {{IEEE International Symposium}} on {{Object Oriented
  Real}}-{{Time Distributed Computing}} ({{ISORC}})}, pages 363--369, 2008.

\bibitem{Kuhne2005a}
Thomas K{\"u}hne.
\newblock What is a {{Model}}?
\newblock In {\em Language {{Engineering}} for {{Model}}-{{Driven Software
  Development}}}, volume 04101. {Internationales Begegnungs- und
  Forschungszentrum f\"ur Informatik (IBFI)}, 2005.

\bibitem{Schramm2014}
Dieter Schramm, Manfred Hiller, and Roberto Bardini.
\newblock {\em Vehicle Dynamics}.
\newblock {Springer}, 2014.

\bibitem{Tao2019}
Fei Tao, He~Zhang, Ang Liu, and A.~Y.~C. Nee.
\newblock Digital {{Twin}} in {{Industry}}: {{State}}-of-the-{{Art}}.
\newblock {\em IEEE Transactions on Industrial Informatics}, 15(4):2405--2415,
  April 2019.

\bibitem{Fuller2020}
Aidan Fuller, Zhong Fan, Charles Day, and Chris Barlow.
\newblock Digital {{Twin}}: {{Enabling Technologies}}, {{Challenges}} and
  {{Open Research}}.
\newblock {\em IEEE Access}, 8:108952--108971, 2020.

\bibitem{Wright2020}
Louise Wright and Stuart Davidson.
\newblock How to tell the difference between a model and a digital twin.
\newblock {\em Advanced Modeling and Simulation in Engineering Sciences},
  7(1):13, December 2020.

\bibitem{Weyns2012}
Danny Weyns, M.~Usman Iftikhar, Didac~Gil {de la Iglesia}, and Tanvir Ahmad.
\newblock A survey of formal methods in self-adaptive systems.
\newblock In {\em Proceedings of the {{Fifth International C}}* {{Conference}}
  on {{Computer Science}} and {{Software Engineering}} - {{C3S2E}} '12}, pages
  67--79, {Montreal, Quebec, Canada}, 2012. {ACM Press}.

\bibitem{Chen2018}
Tao Chen, Rami Bahsoon, and Xin Yao.
\newblock A {{Survey}} and {{Taxonomy}} of {{Self}}-{{Aware}} and
  {{Self}}-{{Adaptive Cloud Autoscaling Systems}}.
\newblock {\em ACM Computing Surveys}, 51(3):1--40, June 2018.

\bibitem{Kephart2003}
J.O. Kephart and D.M. Chess.
\newblock The vision of autonomic computing.
\newblock {\em Computer}, 36(1):41--50, January 2003.

\bibitem{Lasi2014}
Heiner Lasi, Peter Fettke, Hans-Georg Kemper, Thomas Feld, and Michael
  Hoffmann.
\newblock Industry 4.0.
\newblock {\em Business \& information systems engineering}, 6(4):239--242,
  2014.

\bibitem{Bencomo2014}
Nelly Bencomo, Robert France, Betty H.~C. Cheng, and Uwe A{\ss}mann, editors.
\newblock {\em Models@run.Time}, volume 8378 of {\em Lecture {{Notes}} in
  {{Computer Science}}}.
\newblock {Springer International Publishing}, {Cham}, 2014.

\bibitem{Bencomo2019}
Nelly Bencomo, Sebastian G{\"o}tz, and Hui Song.
\newblock Models@run.time: A guided tour of the state of the art and research
  challenges.
\newblock {\em Software \& Systems Modeling}, 18(5):3049--3082, October 2019.

\bibitem{Karimadini2018}
Mohammad Karimadini, Ali Karimoddini, and Abdollah Homaifar.
\newblock A {{Survey}} on {{Fault}}-{{Tolerant Supervisory Control}}.
\newblock In {\em 2018 {{IEEE}} 61st {{International Midwest Symposium}} on
  {{Circuits}} and {{Systems}} ({{MWSCAS}})}, pages 733--738, {Windsor, ON,
  Canada}, August 2018. {IEEE}.

\bibitem{SciPy1.0Contributors2020}
{SciPy 1.0 Contributors}, Pauli Virtanen, Ralf Gommers, Travis~E. Oliphant,
  Matt Haberland, Tyler Reddy, David Cournapeau, Evgeni Burovski, Pearu
  Peterson, Warren Weckesser, Jonathan Bright, St{\'e}fan~J. {van der Walt},
  Matthew Brett, Joshua Wilson, K.~Jarrod Millman, Nikolay Mayorov, Andrew
  R.~J. Nelson, Eric Jones, Robert Kern, Eric Larson, C~J Carey, {\.I}lhan
  Polat, Yu~Feng, Eric~W. Moore, Jake VanderPlas, Denis Laxalde, Josef
  Perktold, Robert Cimrman, Ian Henriksen, E.~A. Quintero, Charles~R. Harris,
  Anne~M. Archibald, Ant{\^o}nio~H. Ribeiro, Fabian Pedregosa, and Paul {van
  Mulbregt}.
\newblock {{SciPy}} 1.0: Fundamental algorithms for scientific computing in
  {{Python}}.
\newblock {\em Nature Methods}, 17(3):261--272, March 2020.

\bibitem{Legaard2020}
Christian~M{\o}ldrup Legaard, Cl{\'a}udio Gomes, Peter~Gorm Larsen, and
  Frederik~F. Foldager.
\newblock Rapid {{Prototyping}} of {{Self}}-{{Adaptive}}-{{Systems}} using
  {{Python Functional Mockup Units}}.
\newblock In {\em 2020 {{Summer Simulation Conference}}}, {{SummerSim}} '20,
  page to appear, {Virtual event}, 2020. {ACM New York, NY, USA}.

\bibitem{Kalman1960}
Rudolph~Emil Kalman.
\newblock A {{New Approach}} to {{Linear Filtering}} and {{Prediction
  Problems}}.
\newblock {\em Journal of Basic Engineering}, 82(1):35, March 1960.

\bibitem{Thule2020}
Casper Thule, Cl{\'a}udio Gomes, and Kenneth Lausdahl.
\newblock Formally {{Verified FMI Enabled Data Broker}}: {{RabbitMQ FMU}}.
\newblock In {\em 2020 {{Summer Simulation Conference}}}, {{SummerSim}} '20,
  page to appear, {Virtual event}, 2020. {ACM New York, NY, USA}.

\end{thebibliography}

\end{document}